\documentclass[twocolumn, amsmath,amssymb,amssymb,aps, prl,showpacs,prl]{revtex4-1}
\usepackage{graphicx,epsfig,epstopdf}
\usepackage{times}
\usepackage{mathrsfs}
\usepackage{dcolumn}
\usepackage{color} 
\usepackage{bm}
\usepackage{braket}
\usepackage{mathrsfs}
\begin{document}
\title{{\Large{Quantum-Coherence-Enhanced Surface Plasmon Amplification by\\
Stimulated Emission of Radiation}}}
\author{Konstantin  E. Dorfman$^{1,2}$}
\email{kdorfman@uci.edu}
\author{Pankaj K. Jha$^{1,3}$}
\email{pkjha@physics.tamu.edu}
\author{Dmitri V. Voronine$^{1,3}$, Patrice Genevet$^{4}$, Federico Capasso$^{4}$ and Marlan O. Scully$^{1,3,5}$}
\affiliation{$^{1}$Texas A$\&$M University, College Station, Texas 77843, USA\\
$^{2}$University of California Irvine, California 92697, USA\\
$^{3}$Princeton University, Princeton, New Jersey 08544, USA\\
$^{4}$Harvard University, Cambridge, Massachusetts 02138, USA\\
$^{5}$Baylor University  Waco, Texas 76798, USA}
\date{\today}
\begin{abstract}
We investigate surface plasmon amplification in a silver nanoparticle coupled to an externally driven three-level gain medium, and show that quantum coherence significantly enhances the generation of surface plasmons. Surface plasmon amplification by stimulated emission of radiation is achieved in the absence of population inversion on the spasing transition, which reduces the pump requirements. The coherent drive allows us to control the dynamics, and holds promise for quantum control of nanoplasmonic devices.
\end{abstract}
\maketitle

Quantum nanoplasmonics is a promising active field of research which involves quantum mechanical control of plasmon resonances \cite{Scholl2012,Esteban2012}, quantum optical applications using plasmons \cite{Chang2006,Kolesov2009} and the development of active plasmonic devices \cite{StockOE}. Surface plasmons (SPs) localize the light within subwavelength volumes which makes them an ideal tool for enhancing and controlling the light-matter interaction at the nanoscale. Although amplification of light is generally bound to the limit of diffraction, it has been shown that stimulated emission of SPs can coherently amplify optical fields in smaller volumes,  thus generating highly localized field of interest for both applied and fundamental physics \cite{StockOE,spaser1,spaserexpl,spaserdet}. Spasers and nano-lasers have been experimentally demonstrated recently \cite{nanolas,Oulton2009,dipolarmedium,Lu12,Ding13} and may find applications, for example, in sensing, bio-imaging and spectroscopy \cite{Hecht2007}. Recent progress in nanophotonics has led to the possibility of controlling many aspects of light with a single layer of nanostructured elements \cite{Yu11, Gen12, Xia13}. These techniques based on optical phase discontinuities, require light amplification at the nanoscale to achieve high throughput.

Several technical challenges, however, must be overcome in order to realize reliable, efficient, high-gain spasers. First, spasers have high thresholds which may limit their use in applications \cite{khurgin2012}. Second, in addition to the threshold problem, spasers have low efficiencies, generating only few plasmons per spasing mode \cite{StockOE,spaser1,spaserexpl,spaserdet}. These limitations of spasers with two-level gain medium are related to the effect of gain saturation caused by the feedback of SP modes on the gain medium. After a short time, absorption and emission become equal, leading to saturation. One way to circumvent this is by adding a saturable absorber (bistable spaser) \cite{spaserdet} which poses technological challenges. Moreover, further increase of the input field intensity leads to extra Ohmic losses due to heating of the metallic surface.

\begin{figure}[t]\label{fig:SWI}
\centerline{\includegraphics[trim=0cm 1cm 0cm 0cm,angle=0, width=0.45\textwidth,angle=0]{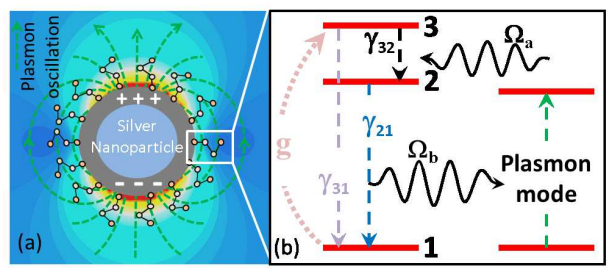}}
 \caption{Schematic of the coherence-enhanced spaser. (a) A silver nanosphere surrounded by three-level quantum emitters such as atoms, molecules, rare earth ions or semiconductor quantum dots placed in the near field of a dipolar plasmon mode. (b) The three-level gain medium is excited by an external incoherent pump $g$ to the upper state $|3\rangle$ which decays to states  $|1\rangle$ and $|2\rangle$. The $|2\rangle \to |1\rangle$ transition is nearly resonant with the plasmon mode of the silver nanosphere such that the state  $|2\rangle$ decays by emitting SPs via energy transfer. The plasmonic oscillations of the nanosphere stimulate this emission, supplying coherent feedback for the spaser. The physical mechanisms involved in the spasing process are analogous to electromagnetically induced transparency and coherent population trapping: an external coherent drive $\Omega_{a}$ is applied to the transition $|2\rangle \to |3\rangle$ creating an asymmetry between absorption and stimulated emission on the spasing transition $|2\rangle\to |1\rangle$, enabling spasing without population inversion on the spasing transition.}
\end{figure}

The field of nanophotonics would benefit from the development of nanoscale coherent sources with increased field intensity output, lower threshold and with performance controlled via external fields. In this Letter, we apply a concept similar to lasing without inversion (LWI) \cite{ScullyZub, Harris, EIT, LWI1, LWI2, Khu96} to SP generation and amplification. Three-level systems experience Fano-type interference in their absorption profile due to two possible absorption paths \cite{ScullyZub}. This Fano-type interference generates an asymmetry between absorption and stimulated emission which may lead to LWI. We consider a three-level gain medium with one of the transitions coupled to a plasmon mode. Using quantum coherence, we mitigate the SP absorption, achieving gain even without population inversion on the spasing transition in the steady-state regime. We theoretically demonstrate that the performance of an externally driven three-level spaser may be enhanced compared to conventional two-level spasers. The introduction of an external driving field that controls the quantum coherence in the gain medium decreases the spasing threshold and increases the number of SPs. We show that coherence-enhanced spasers can be robust against decoherence and that the coherent drive provides a means to control their dynamics.
 
The spasing process with a three-level gain medium is shown in Fig. 1. The gain medium is a generic system of three-level quantum emitters such as atoms, molecules, rare earth ions or semiconductor quantum dots, which have a ground state $|1\rangle$  and two excited states $|2\rangle$ and $|3\rangle$. An incoherent pumping source couples states $|1\rangle$ and $|3\rangle$. Transition $|2\rangle \to |3\rangle$ is driven by an external coherent source  as in the standard LWI theory \cite{ScullyZub}. Transition $|2\rangle \to |1\rangle$ which occurs spontaneously is nearly resonant to the plasmon mode of a metallic nanostructure such as a silver nanosphere and is used to transfer energy from the gain medium to SPs. The local field of SPs in turn provides feedback by stimulating the  $|2\rangle \to |1\rangle$ transition.

In this work, as in standard semi-classical laser theory, we treat the gain medium quantum mechanically, and the SPs and photons classically. The details of the formalism are given in the Supplementary Information. We consider the plasmon and photon annihilation operators $\hat{a}_n$ and $\hat{b}_m$ as complex numbers $a_n$ and $b_m$, respectively, with the time dependence $a_n=a_{0n}e^{-i\nu_s t}$ and $b_m=b_{0m}e^{-i\nu_{23} t}$, where $a_{0n}$ and $b_{0m}$ are slowly varying amplitudes. Therefore, the number of coherent SPs per spasing mode is given by $N_n=|a_{0n}|^2$. Note, that in this model we assume that, unlike the transition $|2\rangle\to|1\rangle$ which is nearly resonant with the plasmons, the transition $|3\rangle\to|2\rangle$ driven by a photon source is not coupled to the metallic nanostructure and does not excite SPs. Such an approximation is valid when the $3\to2$ transition is in a different optical frequency domain, e.g. mid IR relative to the $2\to1$ transition which occurs in visible range. The coherence on the transition $|2\rangle\to|3\rangle$ is maintained by an external source $\Omega_a$ (Fig. 1).

It is more convenient to rewrite the interaction term, in bra-ket notation, with rotating wave approximation as 
\begin{equation}
\begin{split}
 \mathscr{H}_{int}=-\hbar\sum_p\left(\Omega_{b}^{(p)}e^{i\Delta_bt}|2\rangle\langle1|+\Omega_a^{(p)}e^{i\Delta_{a}t}|3\rangle\langle2|+\text{c.c}\right),
\end{split}
\end{equation}
where $\Omega_{b}^{(p)}=-A_n\mathbf{d}_{21}^{(p)}\nabla\phi_n(\mathbf{r}_p)a_{0n}/\hbar$ is the Rabi frequency for the spasing transition $|2\rangle\to|1\rangle$, $\Omega_{a}^{(p)}=-\mathbf{E}_m(\mathbf{r}_p)\mathbf{d}_{23}^{(p)}b_{0m}/\hbar$ is the Rabi frequency for the driving transition $|2\rangle\to|3\rangle$. We define detunings as $\Delta_{b}=\omega_{21}-\nu_{b}$ and $\Delta_{a}=\omega_{32}-\nu_{a}$. Furthermore we assume that the driving field is strong enough so that the number of photons in the mode $m$ is fixed and does not change with time. Thus, $\Omega_{a}^{(p)}$ is a constant, i.e. independent of time.

Introducing $\rho^{(p)}$ as a density matrix of the $p$th gain-medium chromophore, the emitter density matrix elements $\rho_{ij}$ satisfy the Liouville-von Neumann equation
\begin{equation}\label{eq:rho11}
\dot{\rho}^{(p)}=-\frac{i}{\hbar}[\mathscr{H},\rho^{(p)}]-\frac{1}{2}\{\Gamma,\rho^{(p)}\}, 
\end{equation}
where $\{\Gamma,\rho^{(p)}\}= \Gamma\rho^{(p)} +\rho^{(p)}\Gamma$ and the relaxation rates $\Gamma_{21}=\frac{1}{2}(\gamma_{21}+g)+\gamma_{ph}+i\Delta_{b}$, 
$\Gamma_{31}=\frac{1}{2}(\gamma_{31}+\gamma_{32}+g)+\gamma_{ph}+i(\Delta_{a}+\Delta_{b})$, and
$\Gamma_{32}=\frac{1}{2}(\gamma_{31}+\gamma_{21}+\gamma_{32})+\gamma_{ph}+i\Delta_{a}$.
Here $\gamma_{ij}$ are the decay rates for populations, $\gamma_{ph}$is the phase relaxation (or dephasing) rate of the coherence $\rho_{ij}$ \cite{Woggon}, and $g$ is the incoherent pump rate per gain-medium molecule. The SP stimulated emission is described as excitation by the coherent polarization of the gain medium corresponding to the transition $|2\rangle\to|1\rangle$. The corresponding time evolution equation is obtained using the Heisenberg equation of motion for $a_{0_n}$ and adding the SP relaxation rate $\gamma_n$. It has a similar form to the equation for the two-level gain medium \cite{spaserdet}:
\begin{equation}\label{eq:a0n}
\dot{a}_{0_n}=-\Gamma_{n}a_{0_n}+i\sum_p\varrho_{21}^{(p)}\tilde{\Omega}_{b}^{(p)},
\end{equation}
where $\Gamma_{n}=\gamma_n+i\Delta_{n},\, \tilde{\Omega}_{b}^{(p)}=-A_n\mathbf{d}_{21}^{(p)}\nabla\phi_n(\mathbf{r}_p)/\hbar=\Omega_{b}^{(p)}/a_{0_n}$ is a single plasmon Rabi frequency. 
We assume that the Rabi frequencies are the same for all chromophores and omit the index $(p)$ below. The details of the density-matrix equations are given in the supplement material.

To determine the spasing condition, we assume resonant coupling on the driving transition (i.e. $\Delta_{a}=0$), real fields and $\Omega_{b} \ll \Omega_{a},\Gamma_{21}$. In this limit we keep all the terms for the driving field $\Omega_{a}$ and restrict the field $\Omega_{b}$ to the lowest order. The steady-state population inversions are $\bar{n}_{21}=[(g\gamma_{32}-\gamma_{21}\gamma_{31}-\gamma_{21}\gamma_{32})\Gamma_{32}+2(g-\gamma_{21}-\gamma_{31})\Omega^{2}_{a}]\mathcal{M}$ and $\bar{n}_{32}=(\gamma_{21}-\gamma_{32})g\Gamma_{32}\mathcal{M}$, where $\mathcal{M}^{-1}=\Gamma_{32}[g(\gamma_{21}+\gamma_{32})+\gamma_{21}(\gamma_{31}+\gamma_{32})]+2(2g+\gamma_{21}+\gamma_{31})\Omega^{2}_{a}$. 
This yields the necessary condition for the existence of a steady-state solution 
\begin{equation}\label{eq:disp}
\frac{N_p\tilde{\Omega}_b^{2}}{\Gamma_{n}\Gamma_{21}}\left(\bar{n}_{21}+\frac{\Omega^{2}_{a}}{\Gamma_{31}\Gamma_{32}}\bar{n}_{32}\right)-\frac{\Omega^{2}_{a}}{\Gamma_{21}\Gamma_{31}}=1,
\end{equation}
where $N_p$ is the number of quantum emitters. The condition in Eq.(\ref{eq:disp}) which is valid both below and above the spaser threshold, explicitly includes the coherent drive $\Omega_a$, thus allowing control not only of the spasing threshold but also the spasing frequency and dynamics.
In the absence of the drive ($\Omega_{a}=0)$ and for the fast $|3\rangle \to |2\rangle$ population transfer ($\gamma_{32}\gg g$) Eq.(\ref{eq:disp}) reduces to the spasing condition for a two-level system \cite{spaserdet}. The spasing frequency can be found from Eq. (\ref{eq:disp}):
\begin{equation}\label{eq:freqspaser}
\nu_{s}=\frac{\alpha\omega_{21}+\left(\tilde{\Gamma}_{21}\tilde{\Gamma}_{31}+\Omega^{2}_{a}\right)\omega_{n}}{\alpha+\left(\tilde{\Gamma}_{21}\tilde{\Gamma}_{31}+\Omega^{2}_{a}\right)},
\end{equation}
where $ \alpha=\left[\left(N_p\tilde{\Omega}_b^{2}/\Gamma_{32}\tilde{\Gamma}_{31}\right)\bar{n}_{32}-\gamma_{n}/\tilde{\Gamma}_{31}\right]\Omega^{2}_{a}+\gamma_{n}\tilde{\Gamma}_{31}$ and $\tilde{\Gamma}_{ji}$=Re$[\Gamma_{ji}]$. In the absence of the driving field, Eq.(\ref{eq:freqspaser}) reduces to Eq.(11) of ref.~\cite{spaserdet}. Also, in the absence of the driving field, the spasing frequency is given by a combination of the atomic and the plasmonic resonances. Eq.(\ref{eq:freqspaser}) shows that the driving field modifies the spasing frequency, providing a new external control parameter.

Within a certain range of parameters, it is possible to study the problem analytically. In particular, if $\gamma_{31}\ll\gamma_{32}\ll\gamma_{21}$, in the limit of a strong drive, $\Omega_a\gg\gamma_{21},\gamma_n$, in the vicinity of the threshold $g\sim\gamma_{12}$, the number of surface plasmons  is independent of the driving field and is given for $g>\gamma_{21}$ by $N_n\simeq N_p(g-\gamma_{21})/6\gamma_n$.
On the other hand, if the driving is weak, $\Omega_a\ll\gamma_{12},\gamma_n$, then a similar analysis yields $N_n\simeq N_p\gamma_{32}(g-\gamma_{21})/2\gamma_n \gamma_{21}$.
In the weak drive limit, the linear increment of $N_n$ is smaller, because $\gamma_{32}\ll\gamma_{21}$. Therefore, coherent driving can substantially enhance the number of stimulated surface plasmons. 

\begin{figure}[t]
\centerline{\includegraphics[trim=0cm 0.5cm 0cm 0cm,angle=0,width=0.35\textwidth,angle=0]{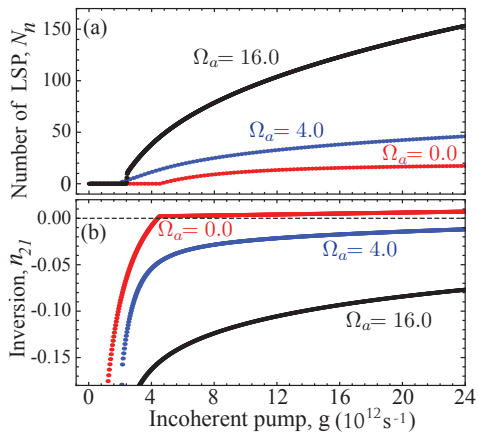}}  
\caption{Steady-state properties of the silver nanosphere spaser with $\hbar\omega_{n}=2.5$ eV as a function of the incoherent pumping rate $g$ for three driving fields: $\Omega_{a}$ = 0 (red), 4$\times$10$^{12}$ s$^{-1}$ (blue) and 16$\times$10$^{12}$ s$^{-1}$ (black). The dephasing rate $\gamma_{ph}$ = 0 . Note that the effect of decoherence are studied in a separate plot in Fig.3.  (a) Number $N_{n}$ of plasmons per spasing mode. Interestingly, $\Omega_{a}$ not only controls the spaser threshold but it also controls the number of generated plasmons. (b) Population inversion $(n_{21})$ on the spasing transition $|2\rangle \leftrightarrow |1\rangle$. Coherent drive allows obtaining a large enhancement of a number of plasmons in the absence of population inversion at the spasing frequency and reduces the spasing threshold.} 
\end{figure} 

Next we present the numerical simulation results obtained by solving the system of equations Eq.(\ref{eq:rho11}). Two types of spasers have been designed based on a spherical core-shell geometry: with the gain medium outside or inside of the nanoparticle \cite{StockOE,spaserexpl}. The first design corresponds to a metal nanosphere surrounded by a shell of gain-medium chromophores, and the second is a metal nanoshell with gain medium inside. Here we consider the first case, i. e. the metal core (nanosphere) surrounded by gain medium molecules, because the nanosphere is the simplest and most thoroughly investigated plasmonic nanostructure with a well-known range of structural and dynamical parameters (see  \cite{Kolwas,Rin13} and Supplementary Information for details). The nanosphere supports localized surface plasmon (LSP) excitations with dipolar, quadrupolar, and higher multipolar modes \cite{Boh83,Pel08}. We consider the lower transition $|2\rangle\leftrightarrow |1\rangle$ of the three-level gain medium coupled to a dipolar SP mode of a silver nanosphere of radius 40 nm and $\hbar\omega_{n}=2.5$ eV. The detuning of the gain medium from the SP mode is $\hbar(\omega_{21}-\omega_{n})$ = 0.002 eV. The external dielectric has the permittivity of $\epsilon_d$ = 2.25. The permittivity of silver was taken from ref. \cite{Johnson1972}. The nanosphere LSP  damping rate was chosen based on the reported values: $\gamma_{n}$ = 5.3$\times$10$^{14}$ s$^{-1}$ with $\Delta_n$ = 3$\times$10$^{12}$ s$^{-1}$ \cite{Kolwas,Nil00,Han09}. 
The number of gain medium chromophores is $N_p\simeq6\times 10^{4}$. The latter has been chosen to match the density of chromophores in \cite{spaser1}. The remaining parameters are listed in the Supplementary Information. Plasmon dephasing may be decreased by varying geometry and coupling to dark resonances \cite{Aes11}. A broad range of structural and dynamic parameters provides flexibility of design and experimental implementation.

Fig. 2(a), obtained by solving the system Eq \ref{eq:rho11} for different values of Rabi frequency $\Omega_{a}$, shows the number of LSPs $N_n$ generated in the presence of the coherent drive $\Omega_{a}$ for the dephasing rate $\gamma_{ph}$ = 0. $N_n$ is enhanced by an order of magnitude for higher values of the pumping rate $g$. This is accompanied by a factor of two decrease of the spasing threshold. Fig. 2(b) shows population inversion on the spasing transition $|2\rangle \leftrightarrow |1\rangle$ for the three choices of drive Rabi frequency $\Omega_{a}$ considered in Fig 2(a). The enhancement of the number of plasmons corresponds to a decrease of the population inversion on the spasing transition. Note, that the negative population inversion on the spasing transition does not mean spasing without population inversion (analogue of lasing without inversion \cite{LWI1,LWI2,Harris}). The overall excited states population is always larger than of the ground state: $\rho_{22}+\rho_{33}>\rho_{11}$ (see supplementary information). The enhancement is due to the quantum coherence in the three-level gain medium generated by the drive $\Omega_{a}$, i. e. breaking of detailed balance on the spasing transition. It is worth noticing two things: first, the coherent drive is applied at a different frequency than both the incoherent pumping or the spasing frequencies; being not resonant with the plasmon mode, we can achieve strong driving field without damaging the nanoparticles. Second, the control of the output can be achieved at low plasmon numbers which may be useful for single plasmon engineering \cite{Kolesov2009} and applications in quantum nanophotonics. To reveal the role of decoherence, we show the number of LSPs $N_n$ versus the incoherent pumping rate $g$ for different dephasing rates $\gamma_{ph}=0, 80, 160$, and $240$$\times$10$^{12}$ s$^{-1}$ in Fig. 3. Even in the case of a  large dephasing rate $\gamma_{ph}=80$$\times$10$^{12}$ s$^{-1}$, which is typical for semiconductor systems, we still observe the effect of enhanced generation of LSPs due to the presence of the coherent drive. Similar effects in atomic systems in the XUV and the optical domains have recently been studied \cite{Jha12,TLWI2, Jha2}.

\begin{figure}[t]
\begin{center}
\centerline{\includegraphics[trim=0cm 2cm 0cm 0cm,angle=0, width=0.35\textwidth,angle=0]{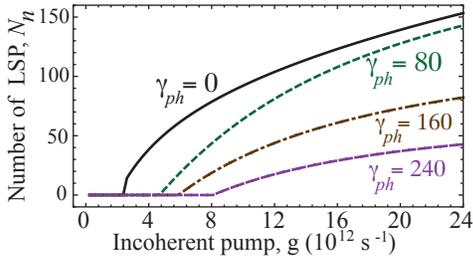}}  
\end{center}
\caption{Number of LSPs $N_n$ versus the incoherent pumping rate $g$ for different dephasing rates $\gamma_{ph}=0, 80, 160$, and $240$$\times$10$^{12}$ s$^{-1}$ for $\Omega_a$ = 16$\times$10$^{12}$ s$^{-1}$. Steady-state coherence-enhanced spasing is robust against decoherence and may be observed even for large dephasing rates.}
\label{scheme}
\end{figure}

\begin{figure}[t]
\begin{center}
\centerline{\includegraphics[trim=0cm 2cm 0cm 0cm,angle=0, width=0.35\textwidth,angle=0]{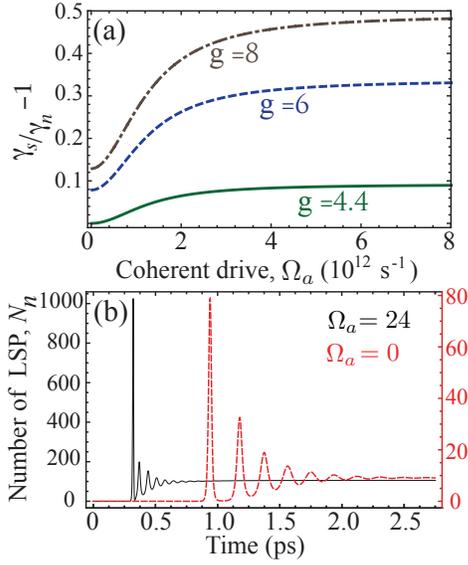}}  
\end{center}
\caption{(a) Ratio of the imaginary part of the Floquet exponent $\gamma_s$ normalized by the SP linewidth $\gamma_n$ for three values of the pumping rate $g$ = 4.4, 6 and 8$\times$10$^{12}$ s$^{-1}$ for $\gamma_{ph}$ = 0. $\gamma_s$ corresponds to the growth rate of the plasmonic field and may be increased by the coherent drive.
(b) Temporal profiles of the plasmonic field intensity for $\Omega_{a}$ = 0 and 24$\times 10^{12}$ s$^{-1}$ show that the steady state is achieved much faster for a stronger drive. This corresponds to a larger value of $\gamma_s$ which can be controlled by the drive as shown in (a).}
\label{scheme}
\end{figure}

In the following, we show that the driving field also affects the dynamical properties of the spaser.
Fig. \ref{scheme}a shows the ratio of the imaginary part of the Floquet exponent $\gamma_s$ normalized by the SP linewidth $\gamma_n$ for three values of the pumping rate $g$ = 4.4, 6 and 8$\times$10$^{12}$ s$^{-1}$ for the dephasing rate $\gamma_{ph}$ = 0. Larger values of  $\gamma_s$  are obtained for larger values of the pump $g$ and moderate values of the drive $\Omega_{a}$. 
Different values of the Floquet exponent correspond to different dynamics and have been used to classify lasers \cite{Wieczorek2005}.  
Fig.  \ref{scheme}b shows two temporal profiles of the plasmonic field intensity for $\Omega_{a}$ = 0 and 24$\times 10^{12}$ s$^{-1}$ for $g$ = 8$\times 10^{12}$ s$^{-1}$. Stronger drive causes the earlier onset of spasing and the faster approach to the steady state. This corresponds to a larger value of $\gamma_s$ which can be controlled by the drive as shown in Fig.  \ref{scheme}a. Such control of the spaser dynamics may be important for ultrafast nanophotonics applications. Although it provides a solution to create local heating sources, decay of LSPs induces heating of the nanoparticles which might be detrimental to spasers. Faster approach to steady-state would also be helpful in the pulsed drive field regime with appropriate repetition rate, avoiding extensive heat which would otherwise eventually melt the nano-particles. 

Coherence-enhanced spasing operation can be experimentally realized by following two different approaches. The first one consists of coating plasmonic nanoparticles or clusters of nanoparticles with self-assembled layers of molecules or quantum dots. Gain and even spasing in such plasmonic nanostructures have previously been reported using a combination of two-level molecules and nanoparticles \cite{nanolas}. Spasing threshold values reported in various experiments confirm that this approach is unpractical for potential applications, in particular due to the large pumping energy required to achieve population inversion \cite{khurgin2012}. Using surface chemistry one can synthesize three-level molecular systems deposited on nanoparticles by e.g. self-assembly. 

The second approach uses very low loss long range propagating surface plasmon polaritons (LRSPPs). These LRSPPs have been successfully amplified by depositing a doped polymer \cite{dipolarmedium} or a dye-doped dielectric on top of the plasmon waveguides \cite{gather2010}. This scheme, where LRSPPs are amplified with surprisingly low noise due to limited mode partition noise\cite{deleon2011}, can be extended to the three-level gain medium by patterning LRSPP waveguides at the interface between three-level quantum wells. 

In summary, we have proposed a new scheme for creating an efficient coherent source of radiation at the subwavelength scale. We showed that the emission properties of spasers may be enhanced using quantum coherence in a three-level quantum emitter based gain medium placed in the near field of a plasmonic nanostructure. We demonstrated a significant spasing enhancement and reduction of the spasing threshold for moderate driving fields. We have also investigated the effect of dephasing and showed that the enhancement is robust under realistic conditions. The driving field acts as an external control parameter to tune the emission properties of the spaser such as threshold, number of generated plasmons and emission frequency.  The driving field may be chosen to interact with the gain medium at a different frequency from that of the spasing transition, avoiding spurious heating effects which could degrade the performance of the device. Coherent control is achieved in a wide range of plasmon output power and persists down to the single plasmon level at low driving field intensity. Because the driving field also controls the spaser dynamics, it may lead to future ultrafast and controllable nanoplasmonic devices. These concepts may be extended to a broader range of quantum optical phenomena and applied to a variety of plasmonic nanostructures. Plasmonic structures of various sizes and shapes can be synthesized to resonantly enhance the applied coherent drive \cite{artif}. Plasmonic analogs of quantum optical effects such as plasmonic EIT \cite{EITth,EITex} and Fano resonances \cite{Fano} may further improve controllability of spaser properties.

We gratefully acknowledge support for this work by National Science Foundation Grants PHY-1241032 (INSPIRE CREATIV) and EEC-0540832 (MIRTHE ERC), and the Robert A. Welch Foundation (A-1261). P. K. Jha is supported by the Herman F. Heep and Minnie Belle Heep Texas A\&M University Endowed Fund held and administered by the Texas A\&M Foundation.


\begin{thebibliography}{99}
  \bibitem{Scholl2012}J. A. Scholl, A. L. Koh, and J. A. Dionne, Nature 483, 421 (2012).
 \bibitem{Esteban2012}R. Esteban et al., Nature Communications 3, 825 (2012).
 \bibitem{Chang2006}D. E. Chang et al. Phys. Rev. Lett. 97, 053002 (2006).
  \bibitem{Kolesov2009}R. Kolesov et al., Nature Physics 5, 470 (2009).
 \bibitem{StockOE}M. I. Stockman, Opt. Express 19, 22029 (2011).
 \bibitem{spaser1}D. J. Bergman, and M. I. Stockman, Phys. Rev. Lett. 90, 027402 (2003).
 \bibitem{spaserexpl}M. I. Stockman, Nature Photonics 2, 327 (2008).
  \bibitem{spaserdet}M. I. Stockman, J. Opt.  12, 024004 (2010).
 \bibitem{nanolas}M. A. Noginov et al., Nature 460, 1110 (2009).
 \bibitem{Oulton2009}R. F. Oulton et al., Nature 461, 629 (2009).
\bibitem{dipolarmedium}I. De Leon, P. Berini, Nature Photonics 4, 382 (2010).
\bibitem{Lu12} Y.J. Lu et al., Science \textbf{337}, 450 (2012).
\bibitem{Ding13} K. Ding et al., Opt. Express \textbf{21}, 4728 (2013).
 \bibitem{Hecht2007}L. Novotny and B. Hecht \textit{Principles of Nano-Optics}, (Cambridge University Press, 2006).
 \bibitem{Yu11} N. Yu et al.,  Science \textbf{334}, 333 (213011).
\bibitem{Gen12} P. Genevet et al.,  App. Phys. Lett. \textbf{100}, 13101 (2012).
\bibitem{Xia13} X. Yin et al., Science, \textbf{339}, 1405 (2013).
 \bibitem{khurgin2012} J. B. Khurgin, and G. Sun, Nanophotonics. 1, 3 (2012).
\bibitem{EIT}S. E. Harris Phys. Today 50, 36 (1997).
 \bibitem{ScullyZub}M.O. Scully and M. S. Zubairy \textit{Quantum Optics}, (Cambridge Press, London 1997).
\bibitem{LWI1}  O. Kocharovskaya, Phys. Rep. 219, 175 (1992).
\bibitem{LWI2} M.O. Scully, S.Y. Zhu, and A. Gavrielides, Phys. Rev. Lett. 62, 2813 (1989).
\bibitem{Harris} S. E. Harris, Phys. Rev. Lett. 62, 1033 (1989).
\bibitem{Khu96} J.B. Khurgin, E. Rosencher, IEEE J. Quant. Elect. \textbf{32}, 1882 (1996).
 \bibitem{Stock2001} M. I. Stockman, S. V. Faleev, and D. J. Bergman, Phys. Rev. Lett. 87, 167401 (2001).
 \bibitem{Woggon}See chapter 5, for discussion on dephasing in semiconductor QDs in U. Woggon, \textit{Optical Properties of Semiconductor Quantum Dots}, (Springer tracts in modern physics, vol. 136,  1958).
 \bibitem{Rin13} E. Ringe et al., Phys. Chem. Chem. Phys. \textbf{15}, 4110 (2013).
 \bibitem{Kolwas}K. Kolwas and A. Derkachova, Opto-Electron. Rev. 18, 429(2010).
 \bibitem{Boh83} C.F. Bohren and D.R. Huffman \textit{Absorption and Scattering Light by Small Particles}, (Wiley, New York, 1983).
 \bibitem{Pel08} M. Pelton et al., Laser \& Photon Rev. \textbf{2}, 136 (2008).
  \bibitem{Johnson1972} P. B. Johnson, and R. W. Christy, Phys. Rev. B 6, 4370 (1972).
   \bibitem{Nil00} N. Nilius et al., Phys. Rev. Lett. \textbf{84}, 3994 (2000).
 \bibitem{Han09} T. Hanke et al., Phys. Rev. Lett. \textbf{103}, 257404 (2009).
 \bibitem{Aes11} M. Aeschlimann et al., Science \textbf{333}, 1723 (2011).


\bibitem{Jha12}P. K. Jha, A. A. Svidzinsky and M. O. Scully, Laser Phys. Lett. 9, 368(2012).
\bibitem{TLWI2} E. A. Sete et al., IEEE J. Sel. Top. Quantum Electron. \textbf{18}, 541 (2012).
\bibitem{Jha2}P. K. Jha, Coherent Optical Phenomena 1, 25(2013)
\bibitem{Wieczorek2005} S. Wieczorek et al., Phys. Rep. 416, 1 (2005).
 \bibitem{gather2010}M. C. Gather et al., Nature Photonics 4, 457 (2010).
\bibitem{deleon2011}I. De Leon, and P. Berini, Phys. Rev. B 83, 081414 (2011).
 \bibitem{artif}H. Wang et al., Acc. Chem. Res. 40, 53 (2007).
 \bibitem{EITth}S. Zhang et al., Phys. Rev. Lett. 101, 047401 (2008).
 \bibitem{EITex}N. Liu et al., Nature Materials, 8, 758 (2009).
 \bibitem{Fano}N. Verellen et al., Nano Lett., 9, 1663 (2009).
 
 
 
 
 
 \end{thebibliography}
\end{document}